\documentclass[aps,prb,twocolumn]{revtex4}

\usepackage{amsmath, amsthm, amssymb}
\usepackage{graphicx,color}

\usepackage[caption=false]{subfig} 

\renewcommand\Im{\operatorname{Im}}

\newcommand{\Lfun}[2]{{\cal L}^{(#1)}_{#2}}
\newcommand{\Tr}[1]{\operatorname{Tr} \left\{ #1 \right\}}

\newcommand{\LU}{{\sc LUMO} }
\newcommand{\HO}{{\sc HOMO} }

\DeclareMathAlphabet{\gcal}{OMS}{cmsy}{m}{n}
\newcommand{\myT}{{\gcal T}}

\graphicspath{{./Figures/}}

\begin{document}


\title{Probing Maxwell's Demon with a Nanoscale Thermometer}

\author{Justin P.\ Bergfield}
\affiliation{Department of Chemistry, Northwestern University, Evanston, IL, 60208, USA}
\email{justin.bergfield@gmail.com}

\author{Shauna M.\ Story}
\author{Robert C.\ Stafford}
\author{Charles A.\ Stafford}
\affiliation{Department of Physics, University of Arizona, 1118 East Fourth Street, Tucson, AZ 85721}

\date{\today}
\begin{abstract}
A precise definition for a quantum electron thermometer is given, as an electron reservoir coupled locally (e.g., by tunneling) to a sample, and
brought into electrical and thermal equilibrium with it.  A realistic model of a scanning thermal microscope with atomic resolution is then developed, where the 
resolution is limited in ultrahigh vacuum by thermal coupling to the electromagnetic environment. 
We show that the
temperatures of individual atomic orbitals or bonds in a conjugated molecule 
with a temperature gradient across it exhibit quantum oscillations,
whose origin can be traced to a realization of Maxwell's demon at
the single-molecule level.    
These oscillations may be understood in terms of the rules of covalence describing bonding in $\pi$-electron systems. 
Fourier's law of heat conduction 
is recovered as the resolution of the temperature probe is reduced, indicating that the macroscopic law emerges as a consequence of coarse graining.
\end{abstract}

\maketitle



Recent advances in thermal microscopy \cite{Kim11,Yu11,Kim12,Fabian12} have opened the door to understanding nonequilibrium thermodynamics at the 
nanoscale.  The nonequilibrium temperature distribution in a quantum system subject to a thermal or electric gradient can now be probed
experimentally, and a number of fundamental questions can be addressed:  Can significant temperature variations occur across individual
atoms or molecules without violating the uncertainty principle? 
How are the electronic and lattice temperatures related in a nanostructure out of thermal equilibrium?  How does the classical
Fourier law of heat conduction emerge\cite{Dubi09b,Dubi09a} from this quantum behavior in the macroscopic limit?

In order to address these questions theoretically, a definition of a nanoscale thermometer that is both realistic and mathematically rigorous is needed.
According to the principles of thermodynamics, 
a thermometer is a small system (probe) with some readily identifiable temperature-dependent
property that can be brought into thermal equilibrium with the system of interest (sample).  Once thermal equilibrium is established, the net heat
current between probe and sample vanishes,\cite{Engquist81,Caso10,Ming10} and the resulting temperature of the probe constitutes a {\em measurement of the sample
temperature}.  High spatial and thermal resolution require that the thermal coupling between probe and sample is local\cite{Kim12} and that the coupling between the probe
and the ambient environment is small,\cite{Kim11} respectively.

It should be emphasized that, out of equilibrium, the temperature distributions of different microscopic
degrees of freedom (e.g., electrons and phonons) do not, in general, coincide, so that one has to distinguish between measurements of the electron
temperature \cite{Engquist81,DiVentra09} and the lattice temperature.\cite{Ming10,Galperin07}  This distinction is particularly acute in the extreme limit of
elastic quantum transport,\cite{Bergfield09} where electron and phonon temperatures are completely decoupled.


\begin{figure}[b]
	\centering
		\includegraphics[width=2.25in]{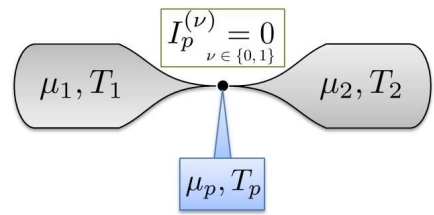}
	\caption{Schematic representation of a temperature probe as the third terminal of a thermoelectric circuit.
}
	\label{fig:probe}
\end{figure}

In this article, we develop a realistic model of a scanning thermal microscope (SThM) operating in the tunneling regime in ultrahigh vacuum,
where the vacuum tunneling gap ensures that phonon heat conduction to the probe is negligible.
Since electrons carry both charge and heat, an additional condition is
necessary to define an electron thermometer.  We proceed by noting that, as a practical matter, in order to reduce the thermal coupling of the thermometer
to the ambient environment, it should form an open electrical circuit (or have very high impedance to ground).  This ensures that, in addition to the 
heat current, the electrical current between sample and probe vanishes.  This is also in accord with the common-sense notion that thermometers are not
current sources or sinks.
An electron thermometer is thus defined as an electron reservoir whose
temperature is fixed by the conditions of electric and thermal equilibrium with the sample:
\begin{equation}
I_p^{(\nu)} =0, \; \nu=0,1,
\label{eq:def_probe}
\end{equation}
where $-eI_p^{(0)}$ and $I_p^{(1)}$ are the electric current and heat current, respectively, flowing into the probe.  
In an ideal measurement, the sample would be the sole source of charge and heat flowing into the probe, but we also consider nonideal measurements, 
where there is an additional thermal coupling to the ambient environment.  In practice, this coupling plays a crucial role in limiting the
resolution of temperature measurements.\cite{Kim11}


In a measurement of the temperature distribution in a conductor subject to thermal and/or electric gradients, the electron thermometer thus serves
as the third terminal in a three-terminal thermoelectric circuit, a generalization of B\"uttiker's voltage probe concept \cite{Buttiker89} (see Fig.\ \ref{fig:probe}).  
Note that the conditions (\ref{eq:def_probe}) allow a local
temperature to be defined under general thermoelectric bias conditions, 
relevant for the analysis of nonequilibrium thermoelectric device 
performance.\cite{Reddy07,Baheti08,Bergfield10,Segalman11}
Previous theoretical analyses of quantum electron thermometers either 
completely neglected thermoelectric effects \cite{Engquist81,Caso10} or considered the measurement scenario (\ref{eq:def_probe}) as only one
of several possibilities,\cite{Sanchez11,Jacquet11}  
while the subtle
definition of local temperature given in Ref.\ \onlinecite{DiVentra09} 
(thermometer causes minimal perturbation of system dynamics) may capture the spirit of our two separate
conditions, at some level.  A recent review of the topic is given in Ref.\ \onlinecite{Dubi11}.


Using our model of a nanoscale electron thermometer, we investigate the nonequilibrium temperature distributions in single-molecule junctions subject to a thermal gradient.
Quantum temperature oscillations analogous to those predicted in one-dimensional systems\cite{DiVentra09}
are predicted in molecular junctions for several different conjugated organic molecules, 
and are explained in terms of the rules of covalence describing bonding in $\pi$-conjugated systems.
In terms of directing the flow of heat, the rules of covalence can be seen as an embodiment of {\em Maxwell's Demon} at the single-molecule level.


It has been argued that in some systems, quantum temperature oscillations can be washed out by either dephasing\cite{Dubi09b} or disorder,\cite{Dubi09a} leading to   
restoration of Fourier's classical law of heat conduction.  However, in molecular junctions the required scattering would be so strong as to dissociate the molecule.
We investigate the effect of finite spatial resolution on the nonequilibrium temperature distribution, and find that Fourier's law emerges naturally as a consequence of
coarse-graining of the measured temperature distribution.  
Thus our resolution of the apparent contradiction between Fourier's macroscopic law of heat conduction and the predicted non-monotonic temperature variations
at the nanoscale is that the quantum temperature oscillations are really there, provided the temperature measurement is carried out with sufficient resolution to
observe them, but that Fourier's law emerges naturally when the resolution of the thermometer is reduced.

This paper is organized as follows:  In Sec.\ \ref{sec:lin_resp}, a general linear-response formula for an electron thermometer is derived.  A realistic model of 
a scanning thermal microscope with sub-nanometer resolution is developed in Sec.\ \ref{sec:SThM}, including a discussion of radiative coupling of the probe to the environment.
Results for the nonequilibrium temperature distributions in single-molecule junctions subject to a thermal gradient are presented in Sec.\ \ref{sec:results}.
A discussion of our conclusions is given in Sec.\ \ref{sec:conclude}.

\section{Electronic Temperature Probe}
\label{sec:lin_resp}

Consider a general system
with $M$ electrical contacts.  Each contact $\alpha$ is connected to a reservoir at temperature $T_\alpha$ 
and electrochemical potential $\mu_\alpha$.  In linear response, the electrical current $-eI^{(0)}_\alpha$  and heat current $I^{(1)}_\alpha$ flowing into reservoir $\alpha$ may be expressed as
\begin{equation}
	I_\alpha^{(\nu)} = \sum_{\beta=1}^M \left[ \Lfun{\nu}{\alpha\beta} (\mu_\beta-\mu_\alpha) 
+ \frac{\Lfun{\nu+1}{\alpha\beta}}{T}(T_\beta-T_\alpha) \right],
	\label{eq:LinearResponse_Currents}
\end{equation}
where 
$\Lfun{\nu}{\alpha\beta}$ is a linear-response coefficient.
Eq.\ (\ref{eq:LinearResponse_Currents}) is a {\em completely general} linear-response formula, and applies to macroscopic systems, mesoscopic 
systems, nanostructures, etc., including electrons, phonons, and all other degrees of freedom, with arbitrary interactions between them. 
For a discussion of this general linear-response formula applied to bulk systems, see Ref.~\citenum{Ziman}.
%

%

\begin{figure}[tb]
	\centering
		\includegraphics[width=2in]{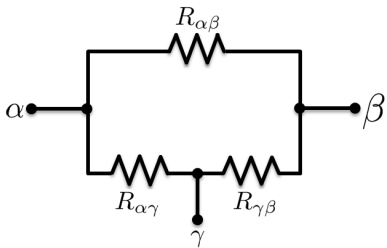}
	\caption{Equivalent circuit for $\tilde{\cal L}_{\alpha\beta}^{(0)}$ and ${\cal L}^{(0)}$, where the resistance 
$R_{\alpha\beta} = [e^2 \Lfun{0}{\alpha\beta}]^{-1}$.  $e^2{\cal L}^{(0)}$ is the loop conductance of the circuit 
and $e^2\tilde{\cal L}_{\alpha\beta}^{(0)}$
is the effective two-terminal conductance between terminals $\alpha$ and $\beta$.
}
	\label{fig:effective_circuit}
\end{figure}

In this article, we consider 
systems driven out of equilibrium by a temperature gradient between reservoirs 1 and 2.  Thermoelectric effects are included, so the chemical potentials
of the various reservoirs may differ.
We consider pure thermal circuits (i.e., open electrical circuits), for which
$I_\alpha^{(0)}=0 \, \forall \,\alpha$.
These conditions may be used to eliminate the chemical potentials $\mu_\alpha$ from Eq.\ (\ref{eq:LinearResponse_Currents}), leading to a
simpler formula for the heat currents
\begin{equation}
I_\alpha^{(1)}=\sum_{\beta=1}^3 \tilde{\kappa}_{\alpha\beta} (T_\beta - T_\alpha).
\label{eq:pure_th}
\end{equation} 
In the absence of an external magnetic field $\Lfun{\nu}{\alpha\beta} =  \Lfun{\nu}{\beta\alpha}$ and the three-terminal thermal conductances 
are given by
\begin{eqnarray}
\tilde{\kappa}_{\alpha\beta} &=& \frac{1}{T}\left[\Lfun{2}{\alpha\beta} 
-\frac{\left[\Lfun{1}{\alpha\beta}\right]^2}{\tilde{\cal L}_{\alpha\beta}^{(0)}}
\right.
\nonumber \\
&-& 
\left.
{\cal L}^{(0)} \!
\left(
\frac{\Lfun{1}{\alpha\gamma}\Lfun{1}{\alpha\beta}}{\Lfun{0}{\alpha\gamma}\Lfun{0}{\alpha\beta}}
+\frac{\Lfun{1}{\gamma\beta}\Lfun{1}{\alpha\beta}}{\Lfun{0}{\gamma\beta}\Lfun{0}{\alpha\beta}}
-\frac{
\Lfun{1}{\alpha\gamma}\Lfun{1}{\gamma\beta}
}{
\Lfun{0}{\alpha\gamma}\Lfun{0}{\gamma\beta}
}
\right)
\right], 
\label{eq:kappatilde}
\end{eqnarray}
with
\begin{equation}
\tilde{\cal L}_{\alpha\beta}^{(0)}=	\Lfun{0}{\alpha\beta}+
\frac{\Lfun{0}{\alpha\gamma}\Lfun{0}{\gamma\beta}}{\Lfun{0}{\alpha\gamma}+\Lfun{0}{\gamma\beta}}
\label{eq:three_term_L0}
\end{equation}
and 
\begin{equation}
\frac{1}{{\cal L}^{(0)}} = \frac{1}{\Lfun{0}{12}} + \frac{1}{\Lfun{0}{13}} + \frac{1}{\Lfun{0}{23}}.
\label{eq:L0_series}
\end{equation}
An equivalent circuit for $\tilde{\cal L}_{\alpha\beta}^{(0)}$ and ${\cal L}^{(0)}$ is given in Fig.\ \ref{fig:effective_circuit}.

The first line of Eq.~(\ref{eq:kappatilde}) resembles the familiar 
two-terminal thermal conductance\cite{Sivan86,Bergfield09b,Bergfield10}  
$\kappa_{\alpha\beta} = \frac{1}{T} \left[ \Lfun{2}{\alpha\beta} - \left(\Lfun{1}{\alpha\beta}\right)^2/\Lfun{0}{\alpha\beta} \right]$,
with $\Lfun{0}{\alpha\beta}$ replaced by Eq.\ (\ref{eq:three_term_L0}).
Since $\Lfun{2}{\alpha\beta}$ is usually the dominant term, $\tilde{\kappa}_{\alpha\beta}$ is often comparable to the two-terminal form
$\kappa_{\alpha\beta}$ (cf. Fig.\ \ref{fig:benzene_para_kappa}).  However, the discrepancy is sizable in some cases.  Although it might be tempting to interpret the second line in Eq.\ (\ref{eq:kappatilde}) as a nonlocal quantum correction to the thermal conductance,
it should be emphasized that this is a generic three-terminal thermoelectric effect, that arises in bulk systems as well as nanostructures.

\subsection{Temperature Measurement}
\label{sec:probe}

In addition to the coupling of the temperature probe to the system of interest, 
we assume 
the probe also has a small thermal coupling $\kappa_{p0}$ to the environment at temperature $T_0$.  The environment could be, for example,
the black-body radiation or gaseous atmosphere surrounding the circuit.  
The heat current flowing from the environment into the probe must be added to Eq.\ (\ref{eq:pure_th}) to determine the total heat current:
\begin{equation}
I_p^{(1)}=\sum_{\beta=1}^2 \tilde{\kappa}_{p\beta} (T_\beta - T_p) + \kappa_{p0}(T_0-T_p).
\label{eq:probe_I1}
\end{equation}
Thermal coupling to the environment is important when the coupling to the system is weak, and is a limiting factor in the 
thermal 
resolution of any temperature
probe.   
The environment is effectively a fourth terminal in the thermoelectric circuit, but since its electrical coupling to the system and probe is zero, 
the thermal conductances $\tilde{\kappa}_{p1}$, $\tilde{\kappa}_{p2}$ in Eq.\ (\ref{eq:probe_I1}) have the three-terminal form (\ref{eq:kappatilde}).  
Solving Eqs.\ (\ref{eq:def_probe}) and (\ref{eq:probe_I1}) for the temperature, we find
\begin{equation}
T_p=\frac{\tilde{\kappa}_{p1} T_1 + \tilde{\kappa}_{p2} T_2 + \kappa_{p0} T_0}{\tilde{\kappa}_{p1} + \tilde{\kappa}_{p2}+  \kappa_{p0}}
\label{eq:Tp}
\end{equation}
for a probe 
in thermal and electrical equilibrium with, and coupled locally to, a system of interest.  
Equations\ (\ref{eq:kappatilde}) and (\ref{eq:Tp}) provide a {\em general definition of an electron thermometer} coupled to a system with a 
temperature gradient across it, in the linear response regime.

\section{Quantum Electron Thermometer}
\label{sec:SThM}

We consider nanoscale junctions 
with weak electron-phonon coupling
operating near room temperature.  Under linear-response conditions,
electron-phonon interactions and inelastic scattering are weak in such systems, and the indirect
phonon contributions to $\Lfun{0}{\alpha\beta}$ and $\Lfun{1}{\alpha\beta}$
can be neglected, while the direct phonon contribution to $\Lfun{2}{p\beta}$ is zero due to the vacuum tunneling gap.  
The linear response coefficients needed to evaluate Eq.\ (\ref{eq:Tp}) may thus be calculated using elastic electron transport theory
\cite{Sivan86,Bergfield09b,Bergfield10}
\begin{equation}
\Lfun{\nu}{\alpha\beta}
= \frac{1}{h} \int dE \; (E-\mu_0)^{\nu}\,\myT_{\alpha\beta}(E) \left(-\frac{\partial f_0}{\partial E}\right),
\label{eq:Lnu}	
\end{equation} 
where $f_0$ is the equilibrium Fermi-Dirac distribution of the electrodes at chemical potential $\mu_0$ and temperature $T_0$.
The transmission function may be expressed as \cite{Datta95,Bergfield09}  
\begin{equation}
{\myT}_{\alpha\beta}(E)={\rm Tr}\left\{ \Gamma^\alpha(E) G(E) \Gamma^\beta(E) G^\dagger(E)\right\},
\label{eq:transmission_prob}
\end{equation} 
where $\Gamma^\alpha(E)$ is the tunneling-width matrix for lead $\alpha$
and $G(E)$ is the retarded Green's function of the junction.  

\subsection{SPM-based temperature probe of a single-molecule junction}

As an 
electron thermometer with 
atomic-scale resolution, we propose using a 
scanning probe microscope (SPM) 
with a conducting tip mounted on an insulating piezo actuator designed to minimize the thermal coupling to the environment.    
The tip could serve e.g.\ as a bolometer or thermocouple,\cite{Kim12} and its temperature could be read out electrically using ultrafine shielded wiring.
The proposed setup is essentially a nanoscale version of the commercially available 
SThM, 
and is analogous to the ground-breaking SThM with 10nm resolution developed by Kim et al.\cite{Kim12}  


Such an atomic-resolution electron thermometer could be used to probe the local temperature distribution in a variety of nanostructures/mesoscopic systems out of equilibrium.  
In the following, we focus on the specific example of a single-molecule junction (SMJ) subject to a temperature gradient, with no electrical current flowing.  
In particular, we consider junctions containing conjugated organic molecules, the relevant electronic states of which are determined by the $\pi$-orbitals.  
We consider transition metal tips, where tunneling is dominated by
the d-like orbitals of the apex atom.\cite{Chen93} 

The tunnel coupling between the tip of the electronic temperature probe and the $\pi$-system of the molecule 
is described by the tunneling-width matrix\cite{Bergfield12a,Chen93} 
\begin{equation}
\Gamma^p_{nm}(E) = 2\pi 
 V_{n} V_m^\ast\, \rho_p(E),
\label{eq:Gamma_p}
\end{equation}
where $n$ and $m$ label $\pi$-orbitals of the molecule, 
$\rho_p(E)$ is the local density of states on
the apex atom of the probe, and $V_m$ is the tunneling matrix element between the evanescent tip wavefunction and orbital $m$ of the molecule.  
Since the temperature probe is in the tunneling regime, and not the contact regime, the phonon contribution to the transport vanishes; 
heat is exchanged between system and probe only via the electron 
tunneling 
characterized by $\Gamma^p$. 

%

\begin{figure}[tb]
	\centering
	\includegraphics{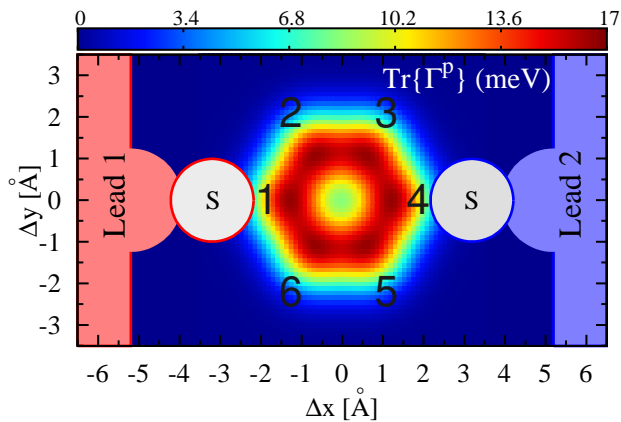} 
	\caption{The calculated spatial map of $\Tr{\Gamma^p}$ for a Pt electron thermometer scanned 3.5\AA\ above the plane of a benzene molecule and a 
schematic representation of a para-benzenedithiol ([1,4]BDT) junction.  Peak values of $\Tr{\Gamma^p}\sim$16.6meV correspond to the centers of the carbon atoms,  
which are numbered in black.  
The sulfur and gold atoms were drawn using their covalent radii of 102pm and 134pm, respectively.  
}
	\label{fig:benzene_para_tr_gamma}
\end{figure}

In Fig.~\ref{fig:benzene_para_tr_gamma}, the trace of $\Gamma^p(E_F)$ is shown  
for a Pt temperature probe held 3.5\AA\ above the plane of a Au-[1,4]benzenedithiol-Au ([1,4]BDT) molecular junction.  
A schematic of the [1,4]BDT junction is also shown, with 
sulfur and gold atoms drawn to scale using their covalent radii of 102pm and 134pm, respectively.  
Peaks in $\Tr{\Gamma^p}$ correspond to the locations of carbon $\pi$-orbitals, labeled with black numbers.  

\begin{figure}[tb]
	\centering
	\includegraphics[width=3in]{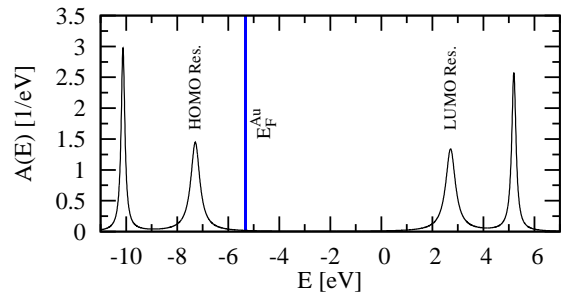} 
\vspace{-0.25cm}
	\caption{Spectral function $A(E)=-1/\pi\Im \Tr{G}$ of a [1,4]BDT (para) junction in the vicinity of the \HO and \LU resonances, calculated using many-body theory including the effect of the the partially charged sulfur atoms on the intramolecular potential.   
In all benzene simulations $\Gamma_{1}=\Gamma_{2}=$0.69eV and the excess charge on the thiol end-groups $\sim$-0.29e, values which give the best agreement with the measured thermopower\cite{Baheti08} and linear-response conductance\cite{Xiao04} measurements.    
The blue line indicates the Fermi energy of gold $E^{\rm Pt}_{\rm F}$=-5.53eV, averaged over the [110], [111], [320], and [331] crystal planes.\cite{CRC}%
}	
	\label{fig:BDT_spectral}
\end{figure}

The density of states (DOS) of the para BDT junction is shown in Fig.~\ref{fig:BDT_spectral}, simulated using our many-body theory including the electrostatic 
influence of the thiol end groups.  The blue vertical line is set at the Pt Fermi energy $E^{\rm Pt}_{\rm F}$=-5.53eV averaged over the [110], [111], [320], and [331] 
crystal planes.\cite{CRC} 
Our many-body theory accurately reproduces the fundamental gap of gas-phase benzene ($\sim$10.4eV) allowing us to unambiguously determine the 
energy-level alignment between the electrodes and molecule. Transport occurs within the HOMO-LUMO gap, but is dominated by the HOMO resonance.   This is true for all of the
molecular junctions considered in this article.

\subsection{Radiative coupling to the environment}
Finally, we assume the coupling of the temperature probe to the ambient environment is predominantly radiative, so that
\begin{equation}
\kappa_{p0} = 4 \epsilon A \sigma T_0^3,
\label{eq:kappa_p0}
\end{equation}
where $\epsilon$ and $A$ are the emissivity and surface area, respectively, of the metal tip, and 
$\sigma=(\pi^2k_B^4/60\hbar^3 c^2)$ is the Stefan-Boltzmann constant.
The tip coupling to the environment should be weak compared to the quantum of thermal conductance 
$\kappa_0 = (\pi^2/3)(k_B^2 T/h)$ in order to resolve quantum effects on the temperature 
within the SMJ.
For a ``spherical tip'' with $A=4\pi R^2$, 
\begin{equation}
\frac{\kappa_{p0}}{\kappa_0} = \frac{8\pi^2 \epsilon}{5} \left(\frac{R k_B T_0}{\hbar c}\right)^2
\stackrel{T_0=300{\rm K}}{=} 0.27 \epsilon \left(\frac{R}{\mu \mbox{m}}\right)^2.
\label{eq:kappa_p0_kappa_0}
\end{equation}
The conducting tip of the temperature probe must thus have linear dimensions $R < 1\mu{\rm m}$ in order to resolve quantum effects at room temperature.
A conducting tip of small volume will also ensure rapid equilibration of the probe with the sample. 

We do not include the direct radiative contribution to $\tilde{\kappa}_{p1}$ and $\tilde{\kappa}_{p2}$.  Since the separation between electrodes 1 and 2 is much less than
the photon thermal wavelength, we consider that black-body radiation from the two electrodes contributes to a common ambient environment at temperature $T_0$.

In all simulations presented here, we consider a Pt electron thermometer with an effective blackbody surface area equal to a 50nm sphere with an emissivity of 0.1,\cite{CRC} 
so that $\kappa_{p0}/\kappa_0 = 6.75 \times 10^{-5}$.
Temperature probes constructed from other metals are predicted to exhibit qualitatively similar effects. 
Eq.\ (\ref{eq:kappa_p0}) may represent an overestimate due to quantum suppression of radiative heat transfer for structures smaller than the thermal wavelength,\cite{Biehs10} in which case the conditions on the probe dimensions would be less restrictive.

\section{Results}
\label{sec:results}

In this section, we investigate the spatial temperature profiles for three Au-benzenedithiol-Au (BDT) junction geometries, a linear [1,6]hexatrienedithiol junction and a poly-cyclic [2,7]pyrenedithiol junction.  The BDT and hexatriene SMJ calculations were performed using a molecular Dyson equation (MDE) many-body transport theory\cite{Bergfield09} in which the molecular $\pi$-system is solved exactly, including all charge and excited states, and the lead-molecule tunneling is treated to infinite order.  The transport calculations for the pyrene junction were performed using H\"uckel theory.
In all cases, the ambient temperature is taken as $T_0$=300K.

%

  \begin{figure}[b]
	\centering
\includegraphics[width=2.7in]{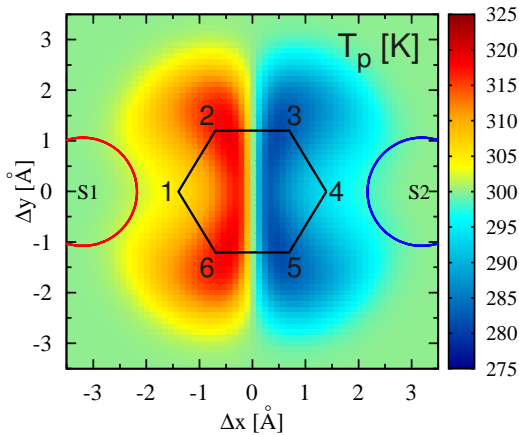}
	\caption{The calculated spatial temperature distribution of a para junction ([1,4]BDT) with $T_1=325K$ and $T_2=275K$, measured by an SThM scanned 
3.5\AA\ above the plane of the carbon nuclei. 
The Pt SThM tip is assumed to be atomically sharp, with an effective blackbody surface area equal to that of a 50nm sphere with an emissivity of 0.1.  
The sulfur linker atoms are indicated schematically by red and blue circles indicating the contacts to the hot and cold leads, respectively.
Quantum oscillations of the temperature are clearly visible in the vicinity of the molecule,  
which can be explained in terms of the Kekul\'e contributing structures shown in Fig.~\ref{fig:kekule}.
}
	\label{fig:benzene_para_tp}
\end{figure} 
 
\subsection{Benzenedithiol junctions}

We investigate the temperature distributions for three Au-benzenedithiol-Au (BDT) junction geometries: the `para' [1,4]BDT junction, 
shown schematically along with the trace of the lead-molecule coupling matrix in Fig.\ \ref{fig:benzene_para_tr_gamma}, the `ortho' [1,2]BDT junction, and the `meta' [1,3]BDT
junction.  
The calculated spatial temperature distribution for a para 
junction is shown in Fig.~\ref{fig:benzene_para_tp}, with $T_1=325K$ and $T_2=275K$.  
The figure illustrates quantum oscillations of the local temperature near the molecule, which are clearly resolvable using our model of a nanoscale electron
thermometer.  
Each of the $\pi$-orbitals of the molecule has a characteristic temperature
different from that of its nearest neighbors: orbitals 2 and 6 
are hot, orbitals 3 and 5 are cold, while orbitals 1 and 4, directly connected to the hot and
cold electrodes, respectively, have intermediate temperatures.  
For weaker thermal coupling of the probe to the ambient environment, thermal oscillations are observed at even larger length scales, while
stronger coupling to the environment reduces resolution of these quantum effects.

Quantum oscillations of the temperature in a nanostructure subject to a temperature gradient are a 
thermal analogue of the voltage oscillations predicted by B\"uttiker \cite{Buttiker89} in a quantum system with an electrical bias. Similar temperature oscillations in 
one-dimensional quantum systems were first predicted by Dubi and Di Ventra.\cite{DiVentra09}

\begin{figure*}[tb]
	\centering
	\subfloat[$\tilde{\kappa}_{p1}$]{
	\includegraphics[width=2.7in]{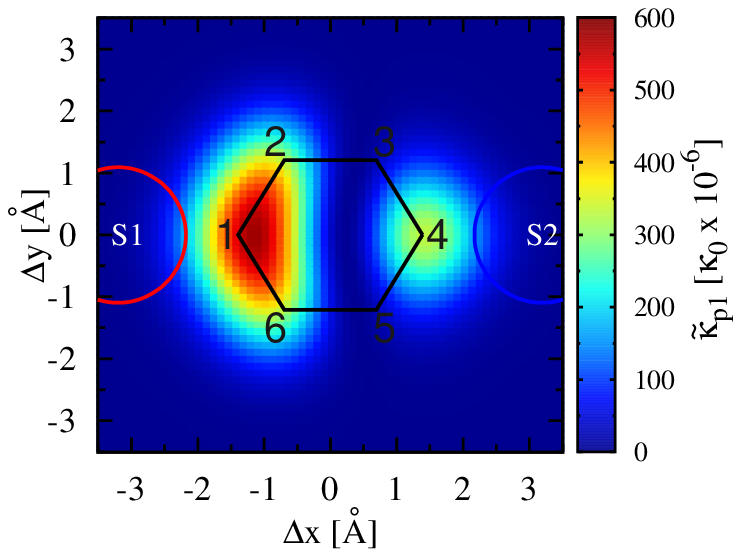}
	 }
	\subfloat[$\tilde{\kappa}_{p1}-\kappa_{p1} $]{
	\includegraphics[width=2.7in]{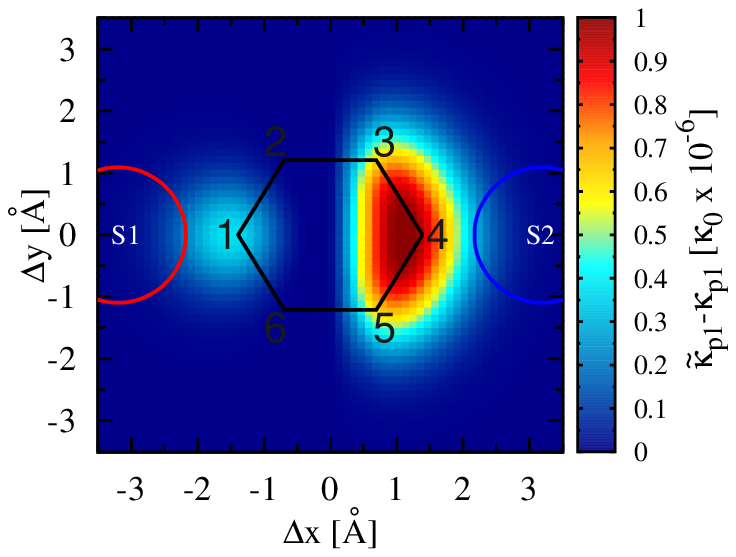}
}%
	 \caption{(a) Thermal conductance $\tilde{\kappa}_{p1}$ between lead $1$ (hot) 
	 and the temperature probe $p$.  $\tilde{\kappa}_{p1}$ is largest 
when $p$ is in an ortho or para configuration relative to the hot electrode, or proximal to it 
(near orbitals 1, 2, 4, or 6), 
and smallest when it is in a meta configuration relative to the hot electrode (near orbitals 3 and 5).  
(b)  The difference between the three- and two-terminal thermal 
conductances $\tilde{\kappa}_{p1}-\kappa_{p1}$, showing the largest errors ($1.03\times 10^{-6} \kappa_0$) occur where $\tilde{\kappa}_{p1}$ is small 
(near sites 3 and 5), indicating large relative errors if a two-terminal formulation were used.  Here $\kappa_0 = (\pi^2/3) k_{\rm B}^2 T_0/h=2.835\times 10^{-4} eV/sK$.  
}
	\label{fig:benzene_para_kappa}
\end{figure*}

In order to understand the temperature oscillations shown in Fig.\ \ref{fig:benzene_para_tp}, it is useful to consider the thermal conductance 
$\tilde{\kappa}_{p1}$ between the probe and the hot electrode. Fig.\ \ref{fig:benzene_para_kappa}(a) indicates that $\tilde{\kappa}_{p\beta}$  
is large when the probe is in the ortho or para configuration relative to electrode $\beta$, as well as when it is proximal to the $\pi$-orbital directly
coupled to the electrode.  However, $\tilde{\kappa}_{p\beta}$ is nearly zero when the probe 
is in the meta configuration relative to electrode $\beta$.  The three-terminal correction to the thermal conductance 
$\Delta \kappa_{p1}$
is plotted in Fig.\ \ref{fig:benzene_para_kappa}(b), which indicates that three-terminal thermoelectric effects lead to a sizable relative correction
to the thermal conductance between the probe and the electrode in the meta configuration, but are otherwise small.

Eq.\ (\ref{eq:Tp}) for the local temperature can be rewritten in the following instructive form:
\begin{equation}
T_p = T_0 + \frac{(\tilde{\kappa}_{p1}-\tilde{\kappa}_{p2}) \Delta T}{\tilde{\kappa}_{p1} + \tilde{\kappa}_{p2} + \kappa_{p0}},
\label{eq:Tp2}
\end{equation}
where $T_1 = T_0 + \Delta T/2$ and $T_{2} = T_0 - \Delta T/2$. 
Thus, when $\tilde{\kappa}_{p1}\gg \tilde{\kappa}_{p2}$, the probe will measure a temperature near $T_1$, and vice versa
when $\tilde{\kappa}_{p1}\ll \tilde{\kappa}_{p2}$, provided the coupling to the environment is 
not too large. On the other hand, the probe will measure a temperature near $T_0$ if $\tilde{\kappa}_{p1}\sim \tilde{\kappa}_{p2}$.
Comparing Figs.\  \ref{fig:benzene_para_tp} and \ref{fig:benzene_para_kappa}, one sees that $\pi$-orbitals 2 and 6 are hot because when the SThM is 
coupled locally to them, it is in an ortho configuration relative to the hot electrode and a meta configuration relative to the cold electrode,
and $\tilde{\kappa}_{\rm ortho}\gg \tilde{\kappa}_{\rm meta}$.  
Orbitals 3 and 5 are cold by symmetry.  On the other hand, orbitals 1 and 4 have
intermediate temperatures since $\tilde{\kappa}_{p1}\sim \tilde{\kappa}_{p2}$ when the SThM is in a para configuration relative to one electrode and proximal to 
the other. 

 \begin{figure}[bt]
	\centering
	\includegraphics[width=2in]{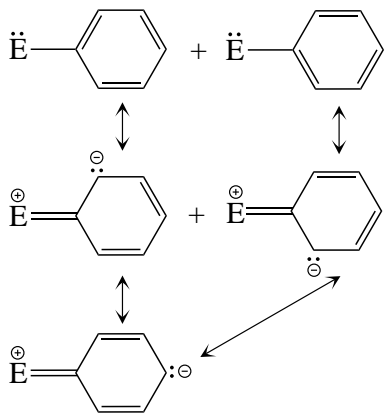}
	\caption{Kekul\'e contributing structures illustrating charge transfer from an electrode E onto benzene.  The rules of covalence in
conjugated systems dictate that electrons from an electrode are available to tunnel onto the temperature probe when it is coupled locally to
the molecule in an ortho or para configuration relative to the electrode.}
	\label{fig:kekule}
\end{figure}

The quantum oscillations in the temperature shown in Fig.\ \ref{fig:benzene_para_tp} can also be understood as consequences of the 
{\em rules of covalence}
in conjugated systems.  Fig.\ \ref{fig:kekule} shows the Kekul\'e contributing structures illustrating charge transfer from an electrode E to a benzene molecule.  
Considering both hot and cold electrodes, the rules of covalence dictate
that electrons from the hot electrode are available to tunnel onto the temperature probe when it is coupled locally to orbitals 2, 4, or 6,
while electrons from the cold electrode are available to tunnel when the probe is near orbitals 1, 3, or 5.  
In addition, electrons from the hot electrode are available to tunnel onto the temperature probe when it is near orbital 1, which is proximal to the hot electrode, while
electrons from the cold electrode are available to tunnel onto the temperature probe when it is near orbital 4, which is proximal to the cold electrode.
Orbitals 2 and 6 thus appear hot, orbitals 3 and 5 appear cold, while orbitals 1 and 4 should exhibit intermediate temperatures by this argument.

\begin{figure}[tb]
	\centering
\includegraphics[width=2.7in]{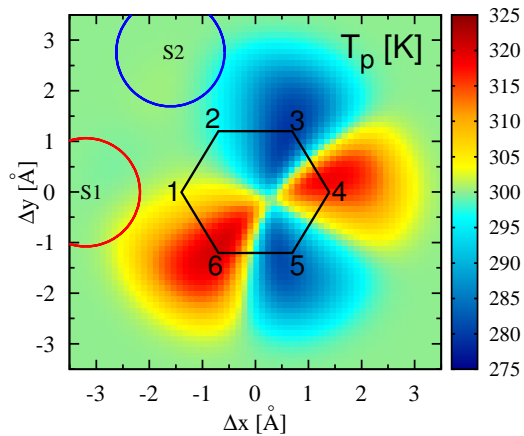}
	\caption{The calculated spatial temperature distribution of an ortho junction ([1,2]BDT) measured under the same conditions described in Fig.\ \ref{fig:benzene_para_tp}.
The quantum temperature oscillations can be explained in terms of the Kekul\'e contributing structures shown in Fig.~\ref{fig:kekule}.}
	\label{fig:benzene_ortho_tp}
\end{figure}

The calculated temperature distribution of an ortho 
junction is shown in Fig.\ \ref{fig:benzene_ortho_tp}, measured under the same conditions discussed in Fig.\ \ref{fig:benzene_para_tp}.
The Kekul\'e contributing structures illustrated in Fig.\ \ref{fig:kekule} dictate that lone pairs from the hot electrode may tunnel to orbitals 2, 4, or 6, 
and lone pairs from the cold electrode may tunnel to orbitals 1, 3, or 5.
Taking into account that orbitals 1 and 2 are also proximal to the linker groups binding the molecule to the hot and cold electrodes,
respectively, and thus exhibit intermediate temperatures, the rules of covalence 
dictate that
orbitals 4 and 6 appear hot, while orbitals 3 and 5 appear cold, in complete agreement with the calculated temperature distribution.

For the para and ortho junctions shown in Figs.\ \ref{fig:benzene_para_tp} and \ref{fig:benzene_ortho_tp},  the rules of covalence act essentially like a {\em Maxwell
demon}, in that they selectively permit electrons from the hot or cold reservoir to tunnel onto the probe when it is at specific locations near 
the molecule, and block electrons from the other reservoir.  
The question might arise whether the actions of this Maxwell demon could lead to a violation of the second law of thermodynamics,
as Maxwell originally hypothesized.  However, in this case there is no violation of the second law, because electrons within the molecule
``remember'' which electrode they came from.  There is no ``mixing'' of the hot and cold electrons in the absence of inelastic scattering,
which is strongly suppressed compared to elastic processes in these junctions at room temperature.



\begin{figure}[tb]
	\centering
\includegraphics[width=2.7in]{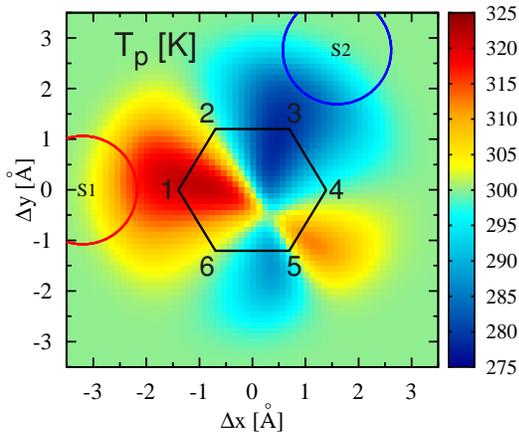}
	\caption{The calculated temperature distribution of a meta junction ([1,3]BDT) measured under the same conditions described in Fig.\ \ref{fig:benzene_para_tp}.
In this case, 
the temperature distribution exhibits well-defined ``bond temperatures'' arising from off-diagonal contributions to the thermal transport,
in addition to the 
``orbital temperatures'' of orbitals 1 and 3 proximal to the hot and cold electrodes, respectively. 
}
	\label{fig:benzene_meta_tp}
\end{figure}

The calculated temperature distribution of a meta BDT junction is shown in Fig.\ \ref{fig:benzene_meta_tp}, measured under the same conditions as in
Figs.\ \ref{fig:benzene_para_tp} and \ref{fig:benzene_ortho_tp}. 
In this case, the temperature distribution is more complicated, exhibiting both well-defined ``orbital temperatures'' (1 and 3) and ``bond temperatures'' (4--5 and
5--6 bonds).  The temperatures of orbitals 1 and 3 can be explained by the arguments given above, while the rules of covalence illustrated in Fig.\ \ref{fig:kekule} indicate
that {\em neither electrons from the hot electrode nor the cold electrode can reach orbital 5}, so that its temperature is indeterminate.
Near orbital 5, off-diagonal contributions to the transmission
dominate due to the suppression of transmission in the meta configuration.  The para transmission amplitude interferes constructively
with the small but nonzero meta transmission amplitude, while the ortho transmission amplitude interferes destructively with the meta transmission
amplitude, so that the 4--5 bond appears hot while the 5--6 bond appears cold.  

\subsection{[1,6]Hexatrienedithiol junction}

\begin{figure}[tb]
	\centering
		\includegraphics{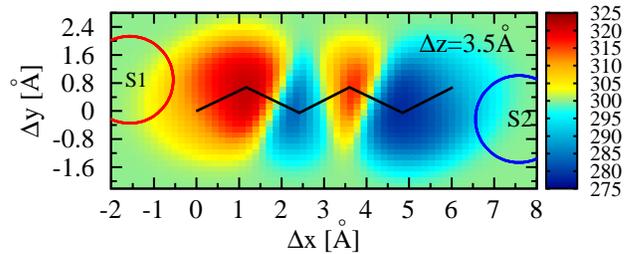}
	\caption{The calculated temperature distribution for a Au-[1,6]hexatrienedithiol-Au molecular junction, measured under the same conditions discussed in
Fig.\ \ref{fig:benzene_para_tp}.
The observed temperature oscillations can be explained using the resonance contributing structures shown in Fig.\ \ref{fig:6site_contributing_structures}.
	}
	\label{fig:hexatriene_tp}
\end{figure}

\begin{figure}[bt]
	\centering
		\includegraphics[width=1.5in]{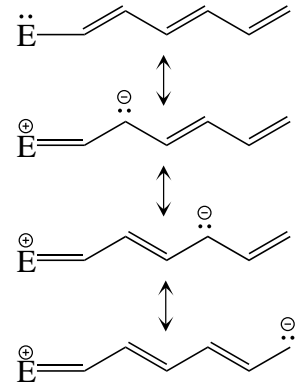}
	\caption{Resonance contributing structures for a hexatriene junction.  For a given electrode $E$, the lone pair can only occupy every other 
$\pi$-orbital, giving rise to the alternating hot, cold temperature profile shown in Fig.~\ref{fig:hexatriene_tp}.}
	\label{fig:6site_contributing_structures}
\end{figure}

The calculated temperature distribution of a [1,6]hexatrienedithiol junction composed of a thioloated 6-site linear molecule (hexatriene) 
covalently bonded to two gold electrodes is shown in Fig.\ \ref{fig:hexatriene_tp}.  
The conditions of the temperature measurement are the same as described in Fig.\ \ref{fig:benzene_para_tp}.
Quasi-one-dimensional temperature oscillations are clearly observable along the length of the molecular wire, consistent with the prediction of Ref.\ \onlinecite{DiVentra09}.
The resonance contributing structures describing electron transfer from an electrode E onto the molecule are shown in Fig.\ \ref{fig:6site_contributing_structures}.
As in the case of the para and ortho-BDT junctions, the rules of covalence are unambiguous, and predict alternating hot and cold temperatures 
for the $\pi$-orbitals along the length of the molecule, with intermediate temperatures for the end orbitals proximal to the two electrodes, 
consistent with the calculated temperature distribution.


\subsection{[2,7]Pyrenedithiol junction}

\begin{figure*}[tb]
	\centering
		\includegraphics[width=6in]{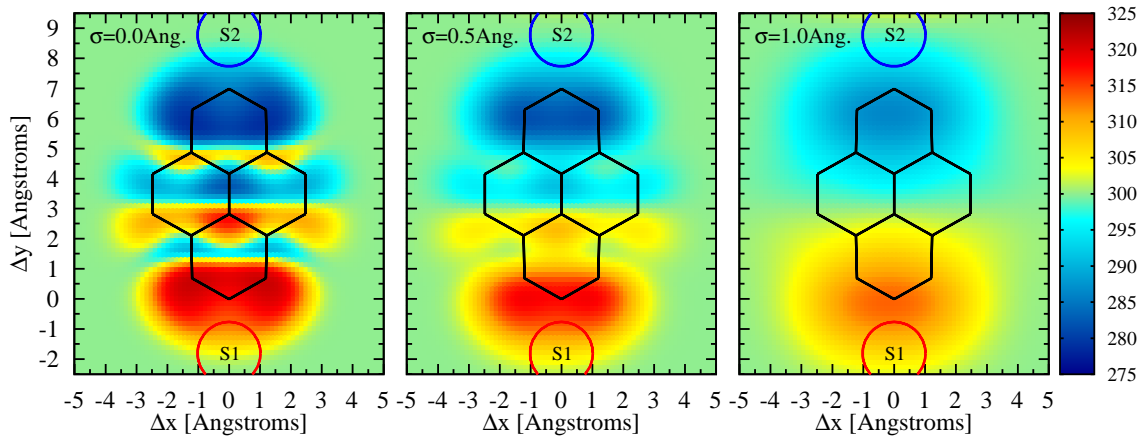}
	\caption{The calculated spatial temperature distribution for a Au-[2,7]pyrenedithiol-Au junction with
$T_1=325K$ and $T_2=275K$, measured at a height of 3.5\AA\ above the plane of the molecule, for three different values of SThM spatial resolution.
The leftmost panel shows the maximum spatial resolution under the specifications of a hypothetical SThM described in Fig.\ \ref{fig:benzene_para_tp}.
The middle and rightmost panels show the measured temperature distributions with reduced spatial resolution obtained by convolving the tip-sample tunnel-coupling $\Gamma^p$
with Gaussian distributions with standard deviations of $\sigma=0.5$\AA\ and 1.0\AA, respectively (full resolution corresponds to $\sigma=0$).
In the rightmost panel, quantum temperature oscillations are no longer resolved, and the temperature distribution resembles a thermal dipole.
Similar results were obtained for a variety of molecular junctions, suggesting that a classical temperature distribution consistent with Fourier's law of heat conduction 
emerges 
due to measurements with limited spatial resolution.
	}
	\label{fig:2_7pyrene_tp}
\end{figure*}

%


As a final example, we consider the effects of finite spatial resolution on the measured temperature distribution of a polycyclic [2,7]pyrenedithiol-Au junction. 
The temperature distribution was calculated for three different values of the SThM spatial resolution in the three panels of Fig.~\ref{fig:2_7pyrene_tp}:  
The leftmost panel shows the maximum spatial resolution under the specifications of a hypothetical SThM described in Fig.\ \ref{fig:benzene_para_tp}.
The middle and rightmost panels show the measured temperature distributions with reduced spatial resolution obtained by convolving the tip-sample tunnel-coupling $\Gamma^p$
with Gaussian distributions with standard deviations of $\sigma=0.5$\AA\ and 1.0\AA, respectively (full resolution corresponds to $\sigma=0$).
Many-body transport calculations for this larger molecule are currently computationally intractable, so we have utilized H\"uckel theory to describe the molecular 
electronic structure, as discussed  
in the Supporting Information.

The left-most panel of Fig.~\ref{fig:2_7pyrene_tp} shows a complex interference pattern of hot and cold regions with a symmetry that mimics the junction itself 
(in this case with two mirror axes).  
More complex molecules such as this highlight the ``proximity effect'' whereby 
the flow of heat from a given electrode to the orbitals in its vicinity is enhanced,
so the molecule is generally warmer near the hot lead and cooler near the cold lead.  
We mention that, although more tedious, the Kekul\'e contributing structures\cite{Camerman65} 
can be used to understand the pattern of temperature variations in this molecule as well. 

Focusing on the middle and right-most panels of Fig.~\ref{fig:2_7pyrene_tp}, 
we find an immediate consequence of the proximity effect as $\sigma$ is increased: 
non-monotonic temperature variations due to quantum interference are washed out, and the underlying temperature gradient appears.
In the rightmost panel, quantum temperature oscillations are no longer resolved, and the temperature distribution resembles a thermal dipole.
Similar results were obtained for a variety of molecular junctions, suggesting that a classical temperature distribution consistent with Fourier's law of heat conduction
emerges when the temperature is measured with limited spatial resolution.


The transition from microscopic quantum temperature oscillations to macroscopic diffusive behavior and Fourier's law is still poorly understood.\cite{Dubi11}  
It has been argued that Fourier's law is recovered in systems with sufficient dephasing\cite{Dubi09b} or disorder.\cite{Dubi09a}
However, we have seen that in conjugated organic molecules, the quantum temperature oscillations are intimately connected to the rules of covalence describing the
$\pi$-bonds of the molecule.  Dephasing (or disorder) sufficient to wash out the temperature oscillations would thus necessarilly sever the $\pi$-bonds and dissociate the
molecule.  Since the molecules studied in this article are stable at room temperature, we know that such strong dephasing cannot be present.  
Thus we predict that quantum temperature oscillations will be observed in molecular junctions if temperature measurements with sufficient spatial resolution
are performed, and that Fourier's law is a consequence of coarse-graining due to finite spatial resolution.

%

%

\section{Conclusions}
\label{sec:conclude}


We have proposed a physically motivated and mathematically rigorous definition of an electron thermometer as an electron reservoir coupled locally to
and in both thermal and electrical equilibrium with the system being measured [cf.\ Eq.\ (\ref{eq:def_probe})].  This definition is valid under general nonequilibrium
conditions with arbitrary thermal and/or electric bias.  Based on this definition, we have developed a realistic model of an atomic-resolution SThM operating in 
the tunneling regime in ultrahigh vacuum, where the 
resolution of temperature measurements is limited by the radiative thermal coupling of the probe to the ambient black-body environment.

We used this model of an atomic-resolution SThM to investigate the nonequilibrium temperature distributions of a variety of single-molecule junctions subject to thermal
gradients. Quantum oscillations of the local temperature that can be observed using a SThM with sufficiently high resolution are predicted.  
We show that in many cases, these quantum temperature oscillations may be understood straightforwardly
in terms of the rules of covalence describing bonding in $\pi$-electron systems.
As such, these oscillations are predicted to be extremely robust, insensitive to dephasing 
or disorder 
that is insufficient to dissociate the molecule.
Instead, we show that such quantum interference effects are washed out if the spatial resolution of the SThM is insufficient to observe them, and that the 
temperature distribution then approaches that expected based on Fourier's classical law of heat conduction.
Message: The temperature oscillations are really there, if you look closely enough!


One may wonder whether it is meaningful to define a temperature that varies significantly from place to place at the atomic scale.  Since temperature is related to 
mean thermal energy, does a variation of temperature on a scale comparable to the de Broglie wavelength not violate the uncertainty principle?  Our answer to such
questions is a pragmatic one: By definition, temperature is that which is measured by a thermometer, and the position of a 
thermometer can certainly be
controlled with subatomic precision using standard scanning probe techniques.
We should also emphasize that our proposed thermometer measures the {\em electron temperature}, which may be largely decoupled from the {\em lattice temperature}
in nanoscale junctions.

Finally, let us return to the theme of the title of this article.
We have shown that in a molecular junction containing a conjugated organic molecule,
the rules of covalence act essentially like a {\em Maxwell
demon}, in that they selectively permit electrons from the hot or cold reservoir to tunnel onto the probe when it is at specific locations near 
the molecule, and block electrons from the other reservoir.  
The question might arise whether the actions of this Maxwell demon could lead to a violation of the second law of thermodynamics,
as Maxwell originally hypothesized.  However, in this case there is no violation of the second law, because electrons within the molecule
``remember'' which electrode they came from.  There is no ``mixing'' of the hot and cold electrons in the absence of inelastic scattering.
And we have argued that dephasing due to inelastic scattering is insufficient to perturb this particular embodiment of Maxwell's demon without dissociating the
molecule itself.


\bibliography{refs}

\end{document}